\documentstyle[prb,twocolumn,aps,rotate,epsf,floats]{revtex}

\newcommand{\tw}{\tilde{\omega}}

\def\rbx#1{\raisebox{-0.3ex}{$\scriptstyle #1$}}

\begin{document}
\setlength{\unitlength}{1cm}
\renewcommand{\arraystretch}{1.4}

\title{Dynamical dimer-dimer correlation functions from exact diagonalization}

\author{Ralph Werner}
\address{Physics Department, University of Wuppertal, D-42097
Wuppertal, Germany and \\ Physics Department, Brookhaven National
Laboratory, Upton, NY 11973-5000} 

\date{\today}

\maketitle

\centerline{Preprint. Typeset using REV\TeX}

\begin{abstract}
A regularization method is presented to deduce dynamic correlation
functions from exact diagonalization calculations. It is applied to
dimer-dimer correlation functions in quantum spin chains relevant for
the description of spin-Peierls systems. Exact results for the XY
model are presented. The analysis draws into doubt that the
dimer-dimer correlation functions show the same scale invariance as
spin-spin correlation functions. The results are applied to describe
the quasi-elastic scattering in CuGeO$_3$.
\end{abstract}

%%%%%%%%%%%%%%%%%%%%%%%%%%%%%%%%%%%%%%%%%%%%%%%%%%%%%%%%%%%%%%%%%%%%%

\section{Introduction}

Dynamical correlations characterize the spectral properties of
physical systems. They are accessible by a multitude of experimental
setups. Even static properties such as correlation lengths extracted
from energy integrating methods such as X-ray scattering also require
the theoretical knowledge of the dynamics of the correlations to allow
for comparison. The access to dynamical correlation functions for
physically relevant systems is usually difficult even in exactly
solvable models.\cite{BetheansatzI,BetheansatzII,BetheansatzIII}
Dynamical spin-spin correlation functions in Heisenberg chains have
been widely studied numerically\cite{Starykh97QMC,FLS97} as well as
analytically.\cite{MTBB81,Schulz86LL,Tsvelik} 

Usually the focus lies on the imaginary part of the correlation
functions. The real and the imaginary parts can be Kramers-Kronig
transformed into each other and thus hold the same information. The
information that can be extracted from finite systems accessible by
exact diagonalization (ED) concerning the thermodynamic limit is
limited. The accuracy of the results is different for different
regions of the spectrum. In the case of finite systems it proves thus
useful to actually calculate both quantities to extrapolate to the
thermodynamic limit. The emphasis of this article lies on the
presentation of a method to accurately extract the real part of the
correlation functions.  

The random-phase-approximation (RPA) approach applied by Cross and
Fisher\cite{CF79} to describe the spin Peierls-transition has been
successfully applied to
CuGeO$_3$.\cite{Gros98CGO,Kluemper99CGO,Werner99CGO,Wer99,Holicki00XY}
To compare with neutron or X-ray scattering data the phononic dynamic
structure factor has to be determined.
\begin{equation} 
S({\bf q},\omega) = -\frac{1}{\pi} \frac{
    \sum_\nu {\rm Im} D_\nu({\bf q},\omega)} 
{1-\exp(-\beta\omega)}
\label{dynamical}
\end{equation}
$\nu$ labels the relevant phonon modes.
The retarded normal coordinate propagator $D_\nu({\bf q},\omega)$
has been calculated including the spin-phonon coupling term in RPA and
depends crucially on the dimer-dimer correlation function $\chi({\bf
q},\omega)$. 
\begin{eqnarray}\label{Dretarded}
D_{\nu}({\bf q},\omega) &=& 
\nonumber\\&&\hspace{-10ex}
\frac{D^{(0)}_{\nu}({\bf q},\omega)
   \left( 1- \chi({\bf q},\omega) 
         \sum\limits_{\nu'\neq\nu} 
                    g_{\nu',{\bf q}}g_{\nu',-{\bf q}}\
                    D^{(0)}_{\nu'}({\bf q},\omega) \right)} 
	{1- \chi({\bf q},\omega) 
         \sum_{\nu'} g_{\nu',{\bf q}}g_{\nu',-{\bf q}}\
                    D^{(0)}_{\nu'}({\bf q},\omega)}
\nonumber\\
\end{eqnarray}
The spin-phonon coupling constants are denoted $g_{\nu',{\bf q}}$
and the unperturbed normal coordinate propagator is
$D^{(0)}_{\nu}({\bf q},\omega)$. Especially the real part of
$\chi({\bf q},\omega)$ is of relevance since it determines dominantly
the poles of the phononic correlation
function.\cite{Gros98CGO,Kluemper99CGO,Wer99} This dependence is the
initial motivation for an accurate determination of the dimer-dimer
correlation function as presented here.

The RPA-calculations predict for CuGeO$_3$ the appearance of spectral
weight in the center of the phononic spectrum as a precursor of the
phase transition.\cite{Gros98CGO,Wer99} I use the numerical
results to discuss the temperature dependent intensity of the
predicted precursor in neutron\cite{BradenHabil,Hirota95CGO} and
X-ray\cite{Pouget96CGO} scattering experiments. Also, the resulting 
description of the phonon hardening is consistent with
experiment.\cite{Braden98CGO}

The relevant spin-phonon coupled Hamiltonian for spin-Peierls systems
consists of three parts $H=H_s+H_p+H_{sp}$. Of relevance here is the
Heisenberg spin-chain Hamiltonian 
\begin{equation}\label{Hs}
H_s=J\ \sum_{\bf l}\ {\bf S}_{\bf l}\cdot{\bf S}_{{\bf l}+\hat{z}}
    + J_2\ \sum_{\bf l}\ {\bf S}_{\bf l}\cdot{\bf S}_{{\bf l}+2\hat{z}}
\end{equation}
with the superexchange integrals $J$ and $J_2$ between nearest-neighbor
(NN) and next-nearest-neighbor (NNN) magnetic ions, respectively,
and spin 1/2 operators ${\bf S}_{\bf l}$ at magnetic ion site ${\bf
l}$ in the three-dimensional lattice. $\hat{z}$ is a unit vector along
the spin-chain direction. For a detailed discussion of the harmonic
phonon part $H_p$ and the spin-phonon coupling term $H_{sp}$ please
refer to Ref.\ \onlinecite{Werner99CGO}.

Defining the Fourier transform of the dimer operators 
\begin{equation}\label{defYq}
Y_{-{\bf q}} := \sum_{\bf l}\ e^{i{\bf q}{\bf R}_{\bf l}}\ 
        {\bf S}_{\bf l}\cdot{\bf S}_{{\bf l}+\hat{z}}\,,
\end{equation}
the dimer-dimer correlation function can be written as
\begin{equation}\label{chi0RPA}
\chi(q_z,i\omega_n) = 
-\frac{1}{N}\int_0^\beta d\tau
     \ {\rm e}^{i\omega_n \tau}\
     \left\langle Y_{\bf q}(\tau) Y_{-\bf q}(0) \right\rangle\
\end{equation}
with Matsubara frequencies $\omega_n=2\pi n/\beta$, inverse
temperature $\beta=1/(k_{\rm B}T)$, and number of unit cells $N$. It
depends only on momenta $q_z$ along the spin chains. It has been
calculated in the analytically continuted form, where $i\omega_n \to
\omega + i\epsilon$ with $\epsilon\to 0$, by Cross and
Fisher\cite{CF79} with bosonization techniques as  
\begin{eqnarray}\label{chiCF}
\chi_{\rbx{\rm CF}}(q_z,\omega)&=& 
\nonumber\\&&\hspace{-9ex}
       \frac{-\chi_{\rbx{0}}(\frac{k_{\rm B}T}{J})}{0.35\,k_{\rm B}T}\ 
	I_1\left[\frac{\omega - v_s |q_z - \frac{\pi}{c}|}
{2\pi k_{\rm B}T}\right] \,
	I_1\left[\frac{\omega + v_s |q_z - \frac{\pi}{c}|}
{2\pi k_{\rm B}T}\right]
\nonumber\\
\end{eqnarray}
with the spin-wave velocity $ v_s$, lattice constant $c$ along the
magnetic chains, and the functions
\begin{equation}
I_1(k)=
\frac{1}{\sqrt{8\pi}}\ \frac{\Gamma(\frac{1}{4}-\frac{1}{2}ik)}
	{\Gamma(\frac{3}{4}-\frac{1}{2}ik)}.
\end{equation}
The result has the general form of spin correlation functions obtained
from conformal field theory.\cite{Tsvelik,Schulz86LL} I set $\hbar=1$
in this paper.

The prefactor $\chi_{\rbx{0}}(k_{\rm B}T/J)$ is assumed constant in
field theory but has been shown by Raupach {\it et al.}\ using density
matrix renormalization group (DMRG) studies to be temperature
dependent in the static case and for $q_z=\pi/c$,\cite{Raupach99CGO}
where $c$ is the lattice constant along the magnetic chains. This wave
vector describes the lattice modulation in the spin-Peierls phase and
is thus the relevant one for this paper. I will now turn to the
question of how far the dynamical properties of the correlation
function are correctly described by the field-theoretical result Eq.\
(\ref{chiCF}).

%%%%%%%%%%%%%%%%%%%%%%%%%%%%%%%%%%%%%%%%%%%%%%%%%%%%%%%%%%%%%%%%%%

\section{Regularization}\label{sectionregular}

The correlation function can be calculated after the diagonalization
of the spin Hamiltonian in the spectral representation since
eigenfunctions $|n\rangle$ and eigenvalues $E_n$ are known. All
numerical results in this paper are obtained using periodic boundary
conditions. Defining the matrix elements
\begin{equation}
V_{nm}=\left\langle n\left| Y_{q_z} \right| m \right\rangle
\end{equation}
and the Boltzmann factor
\begin{equation}
f_{nm}(\beta)=\frac{1}{Z}({\rm e}^{-\beta E_n}-{\rm e}^{-\beta E_m}),
\end{equation}
where $Z$ is the partition function, one can write
\begin{eqnarray}
\label{Rechispectral}
{\rm Re}\,\chi(q_z,\omega) &=& \lim_{\epsilon\to 0}\sum_{m,n} 
\frac{f_{nm}(\beta)\left|V_{nm}\right|^2\! (\omega+E_n-E_m)}
     {(\omega+E_n-E_m)^2+\epsilon^2},
\nonumber\\[-2ex]
\\[1ex]
\label{Imchispectral}
{\rm Im}\,\chi(q_z,\omega) &=& -\pi \sum_{m,n} 
f_{nm}(\beta)\left|V_{nm}\right|^2\ 
     \delta(\omega+E_n-E_m).
\nonumber\\[-2ex]
\end{eqnarray}
Since real and imaginary part of the correlation
function are Kramers-Kronig related via
\begin{equation}\label{KK}
{\rm Im}\,\chi(q_z,\omega) = -\frac{1}{\pi}\, 
    {\rm P}\int_{-\infty}^\infty \frac{{\rm Re}\,\chi(q_z,\omega')}
                               {\omega'-\omega}\, d\omega'\,,
\end{equation}
it suffices in principle to know one of them, or, in other words, both
parts contain the full information on the correlation
function. Since the diagonalization of the Hamiltonian usually is
limited to finite system sizes the information that can be extracted
concerning the thermodynamic limit is limited. In that case it proves
useful to actually calculate both quantities. 
Note that for ${\rm Im}\,\chi(q_z,\omega)$ to have a cut off, i.e., the
correlation function has a finite spectrum, ${\rm
Re}\,\chi(q_z,\omega)$ must change sign as a function of $\omega$.

The spectra of finite systems are discrete and thus the correlation
functions Eqs.\ (\ref{Rechispectral}) and (\ref{Imchispectral})
contain a series of poles as a function of $\omega$ at frequencies
$\omega_{nm} = E_m-E_n$. In order to extract information on the
correlation function in the thermodynamic limit, i.e., $N\to\infty$,
one has to distinguish between well defined quasi-particle spectra and
those where interaction induces transitions between many states. In
the XY case excitations of the non-interacting spin-less fermions form
a well defined continuum in the $N\to\infty$ limit\cite{MTBB81} as
well as the singlet excitations out of the ground state in the
Heisenberg chain.\cite{BetheansatzIII}

\subsection{Imaginary part}

The averaging via binning is obtained by replacing the delta function
in Eq.\ (\ref{Imchispectral}) by a step function with an area of
unity. 
\begin{eqnarray}\label{Imchibins}
{\rm Im}\,\chi(q_z,\tw_j^{\rm inf}<\omega<\tw_j^{\rm sup}) &=& 
\nonumber\\&&\hspace{-26ex}
-\pi\!\! \sum_{m,n}\! \frac{f_{nm}(\beta)\left|V_{nm}\right|^2 
                 [\theta(\omega_{nm}-\tw_j^{\rm inf}) -
                  \theta(\omega_{nm}-\tw_j^{\rm sup})]}
{\tw_j^{\rm sup}-\tw_j^{\rm inf}}
\nonumber\\[-2ex]
\end{eqnarray}
$\theta(x)$ is the Heaviside function. 
For systems with well defined quasi-particle excitation spectra with
spectral lines at frequencies $\tw_j$ as the XY-model or Heisenberg
chains at $T=0$, the appropriate choice is such that the interval
boundaries lie in the middle between those spectral lines: 
\begin{eqnarray}\label{defominf}
\tw_j^{\rm inf} &=& (\tw_{j-1}+\tw_{j})/2\,,
\\ \label{defomsup}
\tw_j^{\rm sup} &=& (\tw_{j}+\tw_{j+1})/2\,.
\end{eqnarray}

Introducing the density of states $\eta^{-1}$ with respect to
Bethe-ansatz quantum numbers it can be
shown\cite{BetheansatzIII,KM00,Michaelprivate} that for systems with 
well defined quasi-particle excitation spectra Eq.\
(\ref{Imchispectral}) can be represented scaled with the system size
as
\begin{equation}\label{Imchidens1}
{\rm Im}\,\chi(q_z,\tw_j) =
-  \lim_{L\to\infty}\frac{L}{2}\,\eta[\tw_j] \sum_{m,n}^{\tw_j=E_n-E_m}
            f_{nm}(\beta)\left|V_{nm}\right|^2 .                 
\end{equation}
The sum covers only values of $n$ and $m$ such that $\tw_j=E_n-E_m$,
$L$ is the number of magnetic ions on a chain. Approximating the
derivative in the density of states for a finite system as 
\begin{equation}\label{eta}
\eta^{-1}[\tw_j]\approx 
\frac{1}{2}\,\frac{\tw_{j}-\tw_{j-1}}{2\pi/L}+
\frac{1}{2}\,\frac{\tw_{j+1}-\tw_{j}}{2\pi/L}
\end{equation}
Eq.\ (\ref{Imchidens1}) becomes
\begin{equation}\label{Imchidens}
{\rm Im}\,\chi(q_z,\tw_j) =
\sum_{m,n}^{\tw_j=E_n-E_m} 
            \frac{-2\pi\ f_{nm}(\beta)\left|V_{nm}\right|^2}
                 {\tw_{j+1}-\tw_{j-1}}.                 
\end{equation}

Given the definitions Eqs.\ (\ref{defominf}) and (\ref{defomsup}) the
expressions (\ref{Imchibins}) and (\ref{Imchidens}) are identical,
except for the latter being only defined at discrete frequencies. Both
approaches suffer the disadvantage of ambiguity concerning the
definition of the interval Eqs.\ (\ref{defominf}) and (\ref{defomsup})
in one case and of the difference quotient Eq.\ (\ref{eta}) in the
other. For larger system sizes,\cite{BetheansatzIII,KM00} where the
density of spectral lines $\tw_j$ is much higher, this ambiguity
becomes irrelevant.  

For Heisenberg chains at finite temperatures the definition of the
density of states $\eta^{-1}$ with respect to
Bethe-ansatz quantum numbers is inappropriate since a large number of
spectral lines are present (see Sec.\ \ref{sectionHeisenberg}). The
binning procedure Eq.\ (\ref{Imchibins}) has to be applied. At higher 
temperatures the interval boundaries $\tw_j^{\rm sup} = \tw_{j+1}^{\rm
inf}$ usually are chosen equidistantly.\cite{FLS97,Starykh97QMC}

\subsection{Real part}

As can be seen from the denominator of Eq.\ (\ref{Rechispectral}) the
real part of $\chi$ is divergent for frequencies $\omega\to
E_m-E_n$. For arbitrarily small but finite $\epsilon$ the term which
leads to the divergence at $\omega\sim\omega_{nm}=E_m-E_n$ vanished
for $\omega=\omega_{nm}$. 
\begin{equation}
0=\frac{(\omega+E_n-E_m)}{(\omega+E_n-E_m)^2+\epsilon^2}
     \bigg|_{\omega=\omega_{nm}} \quad\forall\ \epsilon\neq 0
\end{equation}
It turns out that for systems with well defined quasi-particle
excitation spectra as the XY-model or Heisenberg chains at $T=0$ and
for an accordingly chosen set of frequencies
$\{\tw_j\}=\{\omega_{nm}\}$ this allows a very accurate determination
of the real part of the correlation function from Eq.\
(\ref{Rechispectral}).   

In the case of Heisenberg chains at finite temperatures additional
spectral lines stemming from transitions out of various thermally
activated states are close to each other. This results in lack of
accuracy since the frequencies $\tw_j$ chosen at $T=0$ now lie
uncontrolled on the shoulders of those additional spectral lines. This
problem can be overcome by suppressing all contributions to the
correlation function value in an interval
$E_n-E_m\in[\tw_j\pm\Delta\omega]$. 
\begin{eqnarray}
\label{Rechiregular}
{\rm Re}\,\chi(q_z,\tw_j) &=& 
\nonumber\\&&\hspace{-10ex}
\sum_{m,n} \frac{f_{nm}(\beta)\left|V_{nm}\right|^2}
     {(\tw_j+E_n-E_m)}\ \theta(|E_n-E_m-\tw_j|-\Delta\omega)
\end{eqnarray}

As discussed in the subsequent sections it turns out that for
$\Delta\omega>0.2J$ ${\rm Re}\,\chi(q_z,\tw_j)$ is a smooth
function of $\Delta\omega$. For the Heisenberg chains discussed
${\rm Re}\,\chi(q_z,\tw_j)$ is constant for
$0.06J<\Delta\omega<0.13J$. At intermediate temperatures 
$0.1J<k_{\rm B}T<0.9J$ the method yields inaccurate results for
$\Delta\omega<0.6J$ since the data points are located on shoulders of 
nearby divergences. At higher temperatures the finite size effects
become negligible,\cite{FLS97} at lower temperatures the limiting
value is given by the accuracy with which the $\tw_j$ are
determined. I show that a choice of $\Delta\omega=0.1J$ yields
reliable results in all parameter regimes.  

For higher frequencies the results for the real part of the correlation
functions are free of finite size effects.

%%%%%%%%%%%%%%%%%%%%%%%%%%%%%%%%%%%%%%%%%%%%%%%%%%%%%%%%%%%%%%%%%%%

\section{XY model}

I demonstrate the regularization method and its applicability for an
exactly solvable case, the XY model. Neglecting the exchange between
the $S^z_{\bf l}$ components of the spin system the magnetic
Hamiltonian reduces to 
\begin{equation}\label{HsXY}
H^{\rm XY}_s=J\ \sum_{\bf l}\left( S^x_{\bf l} S^x_{{\bf l}+\hat{z}}
                   +S^y_{\bf l} S^y_{{\bf l}+\hat{z}} \right).
\end{equation}
I have also neglected the next nearest neighbor coupling $J_2$. The
dimer operator becomes
\begin{equation}\label{defYqXY}
Y^{\rm XY}_{-{\bf q}} := \sum_{\bf l}\ e^{i{\bf q}{\bf R}_{\bf l}}\ 
        \left( S^x_{\bf l} S^x_{{\bf l}+\hat{z}}
                   +S^y_{\bf l} S^y_{{\bf l}+\hat{z}} \right).
\end{equation}

\subsection{Exact results}

The spin operators in this model can be transformed to
non-interacting, spin-less fermions via a Jordan-Wigner
transformation.\cite{Fradkin} The correlation function can then be
determined in the thermodynamic limit.\cite{Wer99,Holicki00XY} 
\begin{eqnarray}\label{XYLindhard}
\chi_{\rbx{\rm XY}}(q_z,\omega) &=& 
\nonumber\\&&\hspace{-11ex}
\lim_{\epsilon\to 0 \atop L\to\infty}
\frac{1}{2L}\sum_{k_z} \frac{\left[1 + \cos(2k_z + q_z)c\right]\
    [f_{k_z}-f_{k_z+q_z}]}{\omega+E_{k_z}-E_{{k_z}+{q_z}}+i\epsilon}
\end{eqnarray}
The energy dispersion is given by $E_{k_z}=J\cos(k_z c)$, $f_{k_z}$ are
Fermi distribution functions, the sum covers the first Brillouin zone,
$L$ is the number of magnetic ions in a chain. In general for $L\ge
10^4$ the result is independent of $L$ for all practical purposes.

The imaginary part of Eq.\ (\ref{XYLindhard}) can be given for
$q_z=\pi/c$ in closed form as\cite{YMV96}
\begin{equation}\label{XYIm}
-{\rm Im}\,\chi^{}_{\rbx{\rm XY}}(\pi/c,\omega)\!=\!
	\frac{\sqrt{(2J)^2-\omega^2}}{(2J)^2}
		\tanh\!\left(\!\frac{\beta\omega}{4}\!\right)\!
                         \theta(2J-|\omega|).
\end{equation}

In the limit $T\to 0$ the real part attains a simple form, too.
Defining $x=\omega/(2J)$ one has for $|x|\le 1$
\begin{equation}\label{ReChiXYTlnull}
{\rm Re}\,\chi^{(0)}_{\rbx{\rm XY}}(\pi/c,\omega)=
	\frac{1}{\pi J}-\frac{\sqrt{1-x^2}}{\pi J}\ \ln
	\left|\frac{1+\sqrt{1-x^2}}{x}\right|
\end{equation}
and for $|x|\ge 1$
\begin{equation}\label{ReChiXYThnull}
{\rm Re}\,\chi^{(0)}_{\rbx{\rm XY}}(\pi/c,\omega)=
	\frac{1}{\pi J}-\frac{\sqrt{x^2-1}}{\pi J}\
        \arcsin\left(1/x\right).
\end{equation}

The imaginary part of the correlation function Eq.\ (\ref{XYLindhard})
is depicted for different temperatures by the solid lines in Figs.\
\ref{FigT0XY} and \ref{FigT3XY} (a), the real part in Figs.\
\ref{FigT0XY} and \ref{FigT3XY} (b).

\subsection{Comparing finite and infinite systems}

In finite systems with periodic boundary conditions the allowed wave
vectors are $k_z(j)=2\pi j/L$ for $L/2$ odd and $k_z(j) = \pi (2j+1)/L$
for $L/2$ even with $j\in\{0,\ldots,L-1)$.\cite{LSM61} For $q_z=\pi/c$
at $T=0$ the only spectral components contributing to the
dimer-dimer correlation function Eq.\ (\ref{XYLindhard}) are
fermion-hole excitations with $\tw_j=-2E_{k_z(j)}$ forming a well
defined continuum in the thermodynamic limit and with excitations only
within the half filled fermionic band.\cite{MTBB81,TM85} In the
example of $L=14$ the values to consider are $\tw_j \in {\cal
W}^{(14)}_{\rm XY}$, where ${\cal W}^{(14)}_{\rm XY} = J \{0.445,
1.247, 1.802, 2\}$.

   \begin{figure}[bt]
   \epsfxsize=0.50\textwidth
   \centerline{\epsffile{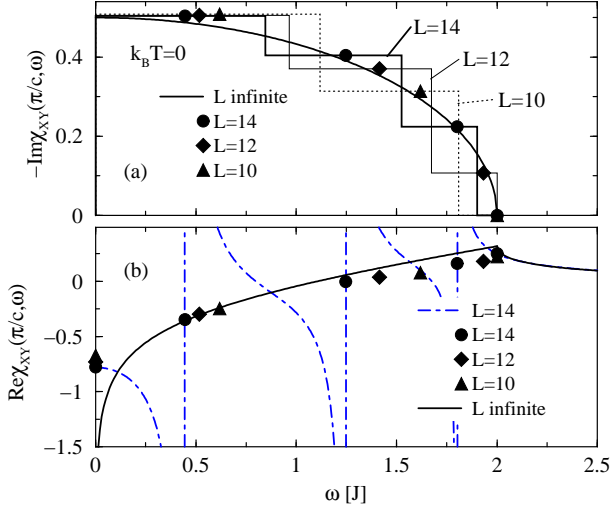}}
   \centerline{\parbox{\textwidth}{\caption{\label{FigT0XY}
   \sl Dimer-dimer correlation function in the XY case at $T=0$. (a)
   imaginary part, (b) real part. Thermodynamic limit results (full
   lines) are obtained via Eqs.\ (\protect\ref{XYIm}) through
   (\protect\ref{ReChiXYThnull}). Step functions are binned as
   given in Eq.\ (\protect\ref{Imchibins}). Symbols are from Eq.\
   (\protect\ref{Imchidens}) (imaginary part) and
   (\protect\ref{XYLindhard}) (real part) at frequencies
   $\tw_j$. Finite size effects of the real part are largest at
   $\omega=0$. Dash-dotted line: exact diagonalization from Eq.\
   (\protect\ref{Rechispectral}).}}}
   \end{figure}

In Fig.\ \ref{FigT0XY} I show different approaches to deduce the
imaginary and real part of the dimer-dimer correlation function from
finite systems compared with the exact result in the thermodynamic
limit ($L\to\infty$, full lines) at $T=0$. The step functions in Fig.\
\ref{FigT0XY} (a) resulting from the binning approach Eq.\
(\ref{Imchibins}) approximate the exact curve quite satisfactorily,
albeit with limited frequency resolution. The scaling approach from
Eq.\ (\ref{Imchidens}) also yields excellent agreement as shown by the
symbols. 

The dash-dotted line in Fig.\ \ref{FigT0XY} (b) is the result of the
real part from Eq.\ (\ref{Rechispectral}) via exact diagonalization
for 14 sites. Above the fermionic spectrum, i.e., $\omega>2J$, the
finite system gives a very accurate result. This is equivalent to the
good agreement of the real time correlation functions from ED and
$L\to\infty$ results for small time scales.\cite{FLS97} The symbols
show the results using Eq.\ (\ref{Rechispectral}) by only regarding
frequencies ${\cal W}^{(L)}_{\rm XY}$ and $\omega=0$. The points give
a satisfactory approximation of the exact curve in the thermodynamic 
limit. The discrepancies due to finite size effects are largest at
$\omega=0$.  

   \begin{figure}[bt]
   \epsfxsize=0.50\textwidth
   \centerline{\epsffile{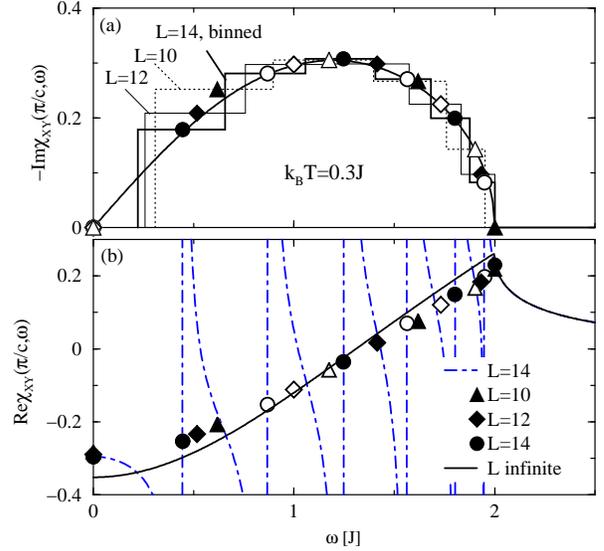}}
   \centerline{\parbox{\textwidth}{\caption{\label{FigT3XY}
   \sl (a) Full line: imaginary part of the correlation function Eq.\
   (\protect\ref{XYIm}) at $k_{\rm B}T=0.3J$. Step functions are
   binned as given in Eq.\ (\protect\ref{Imchibins}). Symbols are from
   Eq.\ (\protect\ref{Imchidens}). (b) Full line: real part of the
   correlation function Eq.\ (\protect\ref{XYLindhard}) at $k_{\rm
   B}T=0.3J$. Dash-dotted line: exact diagonalization from Eq.\
   (\protect\ref{Rechispectral}). Full symbols are values at the
   spectral lines $\tw_j\in {\cal W}^{(L)}_{\rm XY}$ present at $T=0$,
   open symbols are additional spectral lines $\tilde{\cal
   W}^{(L)}_{\rm XY} \setminus {\cal W}^{(L)}_{\rm XY}$ at finite $T$.
   }}}
   \end{figure}

At finite temperatures additional spectral lines are present, namely
for all wave vectors $k_z(j)=\pi j/L$, albeit with temperature
dependent weight. The additional excitations are fermion-hole like
out of thermaly excited superpositions of states with a fermionic
vacancy and an additional fermion 
$|\Phi_\pm\rangle=\sqrt{2}^{-1}(|L/2-1\rangle \pm |L/2+1\rangle)$. For
$L=14$ on has $\tw_j \in \tilde{\cal W}^{(14)}_{\rm XY}$ with
$\tilde{\cal W}^{(14)}_{\rm XY} = J \{0.445, 0.868, 1.247, 1.564,
1.802, 1.950, 2\}$. This circumstance can be used to enhance the
energy resolution of the calculations at finite temperatures. Since
the spectral weight of the additional spectral lines is small at small
temperatures, an application to the imaginary part will only yield
reasonable results for $T\ge0.3J$. Figure \ref{FigT3XY} (a) shows the
excellent agreement with the exact curve (full line) of both the
binned step functions from Eq.\ (\ref{Imchibins}) and the symbols of
the scaling approach Eq.\ (\ref{Imchidens}) using $\tilde{\cal
W}^{(14)}_{\rm XY} \cup \{0\}$. Note that the application of the
scaling approach is only possible here since I can take into account
all spectral lines. $\omega=0$ needs to be included in the spectrum
because of the degeneracy of $|\Phi_+\rangle$ and $|\Phi_-\rangle$.

The calculation with the extended spectrum $\tilde{\cal W}^{(L)}_{\rm
XY}$ can be applied to the real part in principle at any finite
temperature, albeit with some small scattering of the data points at
$k_{\rm B}T < 0.3J$ as shown for the Heisenberg cases in the subsequent
sections. The continuous line in Fig.\ \ref{FigT3XY} (b) shows the
real part of the correlation function Eq.\ (\ref{XYLindhard}) at
$k_{\rm B}T=0.3J$ for the XY model. The dash-dotted line is the result
from Eq.\ (\ref{Rechispectral}) via exact diagonalization for 14 sites
showing the spectral lines $\tw_j$ as divergences. For $\omega>2J$
finite size effects are absent. The full symbols show the results
using Eq.\ (\ref{Rechispectral}) by only regarding frequencies ${\cal
W}^{(L)}_{\rm XY}$ as present at $T=0$, open symbols are added at 
finite temperatures. The points give an excellent approximation of the
exact curve in the thermodynamic limit.

%%%%%%%%%%%%%%%%%%%%%%%%%%%%%%%%%%%%%%%%%%%%%%%%%%%%%%%%%%%%%%%%%%%%%

\section{Heisenberg model}\label{sectionHeisenberg}

I now determine the dimer-dimer correlation function for Heisenberg
chains with the Hamiltonian given in Eq.\ (\ref{Hs}) in the case of
$\alpha=J_2/J=0$.

\subsection{Zero temperature}

At $T=0$ the spectral lines $\tw_j \in {\cal W}^{(L)}_{0}$ at
$q_z=\pi/c$ can be identified as the magnetization conserving
(singlet) excitations out of the ground state determined via the Bethe
ansatz, which form a well defined continuum in the thermodynamic
limit.\cite{BetheansatzI,BetheansatzII} Equation (\ref{Imchidens}) can
thus be applied and Eq.\ (\ref{Rechispectral}) can be regularized as
discussed in Sec.\ \ref{sectionregular}. 

\begin{figure}[bt]
   \epsfxsize=0.50\textwidth
   \centerline{\epsffile{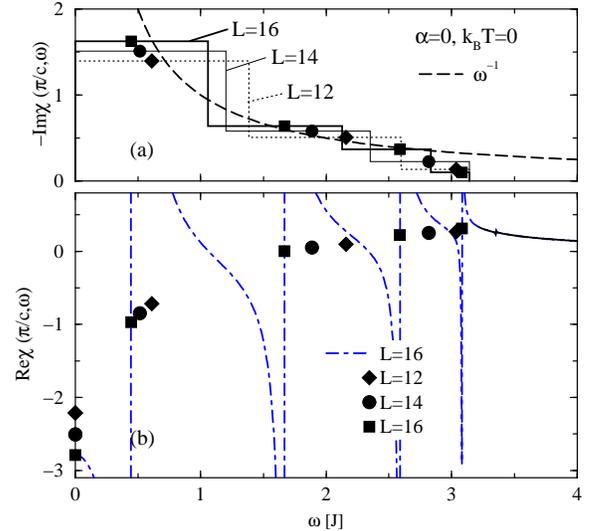}}
   \centerline{\parbox{\textwidth}{\caption{\label{FigT0a00}
   \sl Imaginary part (a) and real part (b) of the dimer-dimer
   correlation function in the unfrustrated Heisenberg chain at
   $T=0$. The imaginary part for finite systems is binned (Eq.\
   (\protect\ref{Imchibins})), each bin holds one spectral line, symbols
   are from Eq.\ (\protect\ref{Imchidens}).  The dashed line is the field
   theoretical result (cf.\ Eq.\ (\protect\ref{ImChiTnull})). The
   symbols for the real part as well as the dash-dotted line are
   obtained using Eq.\ (\protect\ref{Rechispectral}), for
   $\omega>3.14J$ the numerical result is in the thermodynamic limit
   (full line). 
   }}}
   \end{figure}

For a 16 site chain the four spectral lines at ${\cal W}^{(16)}_{0} =
J\{0.446, 1.669, 2.588, 3.083\}$ lead to the imaginary part of the
dimer-dimer correlation function as shown in Fig.\ \ref{FigT0a00}
(a). The bins are obtained using Eq.\ (\ref{Imchibins}) and the
symbols are given using Eq.\ (\ref{Imchidens}). The field theoretical
prediction Eq.\ (\ref{chiCF}) reduces for $T\to 0$
to\cite{MTBB81,Schulz86LL,Tsvelik}    
\begin{equation}\label{ImChiTnull}
-{\rm Im}\,\chi^{(0)}_{\rbx{\rm CF}}(\pi/c,\omega)=
              \frac{C}{\omega}\ 
                         \theta(\Lambda-|\omega|)\,.
\end{equation}
From the fit (dashed line in Fig.\ \ref{FigT0a00} (a)) I conclude
$C\approx 1$. The correspondece of the fit is not too good and
suggests the presence of correction terms. The cut off parameter
$\Lambda\approx\pi J$ is determined via the real part (see below) and
indicates that the spectral weight lies mostly within the two-spinon 
continuum.\cite{dCP62,KM00}

The real part of the dimer-dimer correlation function is shown in
Fig.\ \ref{FigT0a00} (b). For $\omega > \pi J$ the numerical results
are in the thermodynamic limit showing the cut off to be
$\Lambda=\pi J$. For $\omega=0$ finite size effects are significant
since one expects from field theory and by analogy to the XY 
model ${\rm Re}\,\chi^{(0)}_{\rbx{}}(\pi/c,\omega=0) \to -\infty$. For
$\omega\neq 0$ the field theoretical result obtained by the
Kramers-Kronig transform of Eq.\ (\ref{ImChiTnull}) is 
\begin{equation}\label{ReChiTnull}
{\rm Re}\,\chi^{(0)}_{\rbx{\rm CF}}(\pi/c,\omega\neq 0)=
           \frac{C}{\pi\omega}\ 
           \ln\left|\frac{\Lambda+\omega}{\Lambda-\omega}\right|\
           > 0
\end{equation}
with a non-continuous jump to ${\rm Re}\,\chi^{(0)}_{\rbx{\rm
CF}}(\pi/c,0)=-\infty$. In contrast to that for $\omega < J$ the
numerical data in Fig.\ \ref{FigT0a00} (b) indicate ${\rm
Re}\,\chi^{(0)}_{\rbx{}}(\pi/c,\omega) < {\rm
Re}\,\chi^{(0)}_{\rbx{\rm XY}}(\pi/c,\omega) < 0$. This suggests
significant corrections to the field-theoretical results Eq.\
(\ref{ReChiTnull}) and (\ref{ImChiTnull}).

\subsection{Finite temperatures}

\begin{figure}[bt]
   \epsfxsize=0.50\textwidth
   \centerline{\epsffile{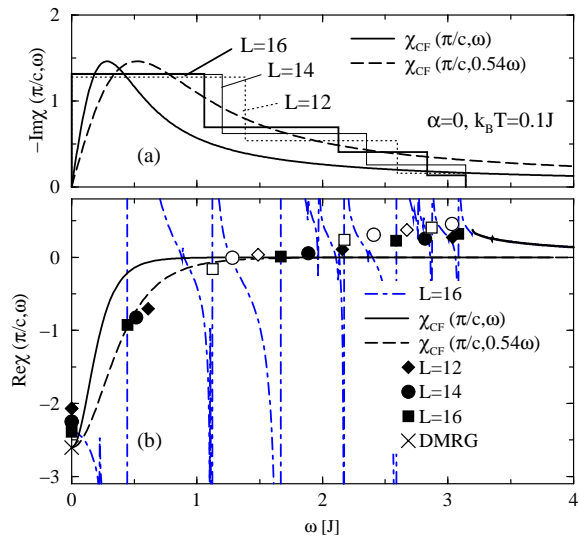}}
   \centerline{\parbox{\textwidth}{\caption{\label{Figure00}
   \sl Imaginary part (a) and real part (b) of the dimer-dimer
   correlation function in the unfrustrated Heisenberg chain at
   $k_{\rm B}T=0.1J$. The imaginary part for finite systems is binned
   as in Fig\ \protect\ref{FigT0a00}. The real part for $\omega<3.14J$
   is obtained using $\Delta\omega=0.1J$ and is given by full symbols
   at frequencies as for $T=0$ and by open symbols at additional
   spectral lines appearing for $T>0$. For $\omega>3.14J$ the
   numerical result is in the thermodynamic limit (thick full
   line). The full lines are the Cross and Fisher result Eq.\
   (\protect\ref{chiCF}), the dashed lines are rescaled according to
   Eq.\ (\protect\ref{Chiscale}).
   }}}
   \end{figure}

At finite temperatures additional spectral lines attain significant 
weight as can be seen by the divergences in the real part of the
correlation function Eq.\ (\ref{Rechispectral}) as plotted for
$k_{\rm B}T=0.1J$ in Fig.\ \ref{Figure00} (b) (dash-dotted line with
L=16). The most dominant spectral lines are $\tilde{\cal W}^{(16)}_{0}
= J\{0.446, 1.125, 1.669, 2.174, 2.588, 2.870, 3.083\}$.

The imaginary part of the dimer-dimer correlation function is
determined via the same binning procedure as for $T=0$ and shown in
Fig.\ \ref{Figure00} (a). At $k_{\rm B}T=0.1J$ the weight of the
additional spectral lines $\tilde{\cal W}^{(16)}_{0} \setminus {\cal
W}^{(16)}_{0}$ is too small to improve the frequency resolution as has
been done in the XY case for $k_{\rm B}T=0.3J$. The scaling
regularization Eq.\ (\ref{Imchidens}) is inappropriate at finite
temperatures in the Heisenberg model as discussed in Sec.\
\ref{sectionregular}. 

The real part of the correlation function has to be regularized
according to Eq.\ (\ref{Rechiregular}). The resulting values for
$\Delta\omega=0.1J$ are shown by the symbols in Fig.\ \ref{Figure00}
(b). Full symbols are evaluated at the $\tw_j\in {\cal W}^{(16)}_{0}$, 
open symbols are determined at additional dominant spectral lines and
are less accurate. The finite size effects are of the order of 10\% at
$\omega=0$ as can be seen by comparison with DMRG results in the
thermodynamic limit\cite{Raupach99CGO} shown by the cross in Fig.\
\ref{Figure00} (b). They are negligible for $\omega>\pi J$.

\begin{figure}[bt]
   \epsfxsize=0.50\textwidth
   \centerline{\epsffile{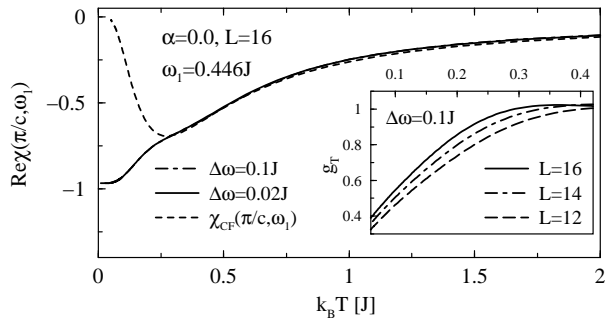}}
   \centerline{\parbox{\textwidth}{\caption{\label{ChivonT00}
   \sl Temperature dependence of ${\rm Re}\,\chi(\pi/c,\omega_1)$ for 
   $\omega_1=0.446J$: the result from Eq.\ (\ref{Rechiregular}) is
   practically independent of $\Delta\omega$, the lines for
   $\Delta\omega=0.02J$ and $\Delta\omega=0.1J$ fall on top of each
   other. The dashed line is the field-theoretical result
   ${\rm Re}\,\chi_{\protect\rbx{\rm CF}}(q_z,\omega_1)$ from Eq.\ 
   (\ref{chiCF}). The inset shows the scaling function defined in Eq.\
   (\ref{Chiscale}) for different chain lengths. The frequency
   corresponding to $L=14$ is $\omega_1=0.516J$, for $L=12$ it is
   $\omega_1=0.61J$.}}}
   \end{figure}

The full lines in Fig.\ \ref{Figure00} (a) and (b) show the
field-theoretical result $\chi_{\rbx{\rm CF}}(q_z,\omega)$ as given by
Eq.\ (\ref{chiCF}). The functions are scaled by $\chi_{\rbx{0}}(k_{\rm
B}T/J)$ to match the $\omega=0$ values obtained by
DMRG.\cite{Raupach99CGO} Both real and imaginary part fit the numerical
data better when rescaling the frequency dependence as $\chi_{\rbx{\rm
CF}}(q_z,\omega\to0.54\omega)$ (dashed lines in Fig.\ \ref{Figure00}
(a) and (b)). The temperature dependence of the discrepancy between
field-theory (dashed line) and numerical result for the real part
(full line) at $\omega=\omega_1=0.446J$ is shown in Fig.\
\ref{ChivonT00}.

As demonstrated in Fig.\ \ref{ChivonT00}, the numerical result for
${\rm Re}\,\chi(\pi/c,\omega_1)$ with $\omega_1=0.446J$ is practically
independent of $\Delta\omega$ for all temperatures. I use this
circumstance to determine the scaling function $g_T$ such that  
\begin{equation}\label{Chiscale}
{\rm Re}\,\chi_{\rbx{\rm CF}}(\pi/c,g_T\,\omega_1)
                      ={\rm Re}\,\chi(\pi/c,\omega_1)\,.
\end{equation}
This procedure shows little finite size effects since the finite size
scaling displaces the points at $\omega_1$ along the slope of the fit
function as can be seen in Fig.\ \ref{Figure00} (b). The resulting
$g_T$ is shown in the inset of Fig.\ \ref{ChivonT00}. The results for
different chain lengths show that $L=16$ yields an upper bound for
$g_T$, the finite size effects decrease with decreasing temperature.

The following arguments show the validity of the real-part results
even for finite regularization parameters $\Delta\omega$:

(i) All symbols shown in Fig.\ \ref{Figure00} (b) follow a smooth
line as expected by comparing with field-theoretical results and the
XY model. The real part changes sign as a function of frequency as
expected. 

(ii) The dependence of ${\rm Re}\,\chi^{(0)}_{\rbx{}}(\pi/c,\tw_j)$ on
$\Delta\omega$ is small. This is shown for $\tw_j=\omega_1=0.446J$ in
Fig.\ \ref{ChivonT00}, where the the values of the correlation
function for $\Delta\omega=0.02J$ and $\Delta\omega=0.1J$ fall on top
of each other for all temperatures.

(iii) The scaling function $g_T$ determined by ${\rm
Re}\,\chi^{}_{\rbx{}}(\pi/c,\omega_1)$ also yields a better 
field-theoretical fit of the imaginary part of the correlation
function.

\section{Frustrated Heisenberg chain}\label{sectionFrustrated}

For $J_2\neq 0$ the Heisenberg model is not integrable any more and
thus the classification of the excitation spectrum at $T=0$ is not
possible from the Bethe ansatz. I discuss here a value of
$\alpha=J_2/J=0.24$ which is among the proposed ones for
CuGeO$_3$\cite{Werner99CGO,Brenig97CGO} and is close the quantum 
critical point\cite{Chitra95DMRG} making the field-theoretical result
given in Eq.\ (\ref{chiCF}) eligible for comparison.

\subsection{Zero temperature}

\begin{figure}[bt]
   \epsfxsize=0.50\textwidth
   \centerline{\epsffile{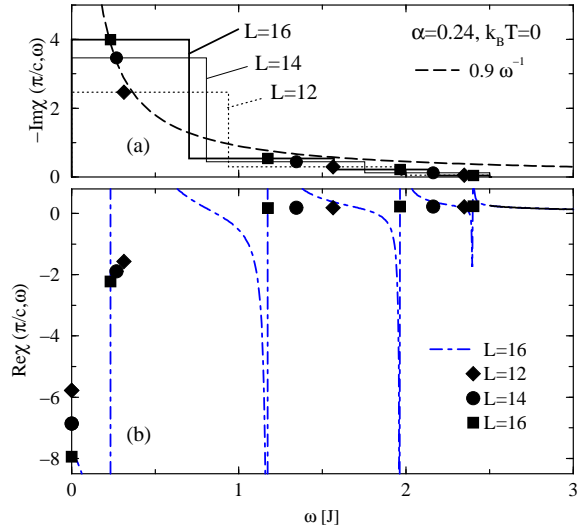}}
   \centerline{\parbox{\textwidth}{\caption{\label{FigT0a24}
   \sl Imaginary part (a) and real part (b) of the dimer-dimer
   correlation function in the frustrated Heisenberg chain at $T=0$
   with $\alpha=0.24$. The imaginary part for finite systems is binned
   (Eq.\ (\protect\ref{Imchibins})), each bin hold one spectral line,
   symbols are from Eq.\ (\protect\ref{Imchidens}).  The dashed line
   is the field-theoretical result (cf.\ Eq.\
   (\protect\ref{ImChiTnull})). The symbols for the real part as well
   as the dash-dotted line are obtained using Eq.\
   (\protect\ref{Rechispectral}), for $\omega>2.5J$ the numerical
   result is in the thermodynamic limit (full line). 
   }}}
   \end{figure}

The dash-dotted line in Fig.\ \ref{FigT0a24} (b) shows that at $T=0$
for a 16 site chain there are four spectral lines $\tw_j \in {\cal
W}^{(16)}_{.24} = J\{0.233, 1.174, 1.963, 2.403\}$ which, by analogy to
the unfrustrated case, may also be identified as the singlet
excitations out of the ground state. It is thus reasonable to suppose
them to form a well defined continuum in the thermodynamic limit and
thus Eqs.\ (\ref{Imchidens}) and (\ref{Rechispectral}) can be applied
($\Delta\omega=0$). Note that $|{\cal W}^{(L)}_{.24}|=|{\cal
W}^{(L)}_{0}|=[L-L{\rm mod}(4)]/4$ for the chain lengths investigated.

The imaginary part of the dimer-dimer correlation function is shown in
Fig.\ \ref{FigT0a00} (a). The bins are obtained using Eq.\
(\ref{Imchibins}) and the symbols are given using Eq.\
(\ref{Imchidens}). The fit to the field theoretical prediction Eq.\
(\ref{ImChiTnull}) (dashed line in Fig.\ \ref{FigT0a24} (a)) leads to
conclude $C = 0.9$. The cut off parameter $\Lambda\approx 2.5J$
determined via the real part (see below) indicates that the
spectral weight lies mostly within a bounded continuum just as in the
unfrustrated case.\cite{Cutoffquote}

The real part of the dimer-dimer correlation function is shown in
Fig.\ \ref{FigT0a24} (b). For $\omega > 2.5 J$ the numerical results
are in the thermodynamic limit suggesting the cut off to be
$\Lambda=2.5J$. For $\omega=0$ finite size effects
are significant since one expects from field theory and by analogy to
the XY model ${\rm Re}\,\chi^{(0)}_{\rbx{}}(\pi/c,\omega=0) \to -\infty$. 

The agreement with the field-theoretical results Eqs.\
(\ref{ImChiTnull}) and (\ref{ReChiTnull}) is better than in the
unfrustrated case but the analogy to the XY model again suggests ${\rm
Re}\,\chi^{(0)}_{\rbx{}}(\pi/c,0.233J) < 0$. It cannot be entirely
excluded though that the discrepancy is due to finite size effects.

\subsection{Finite temperatures}

\begin{figure}[bt]
   \epsfxsize=0.50\textwidth
   \centerline{\epsffile{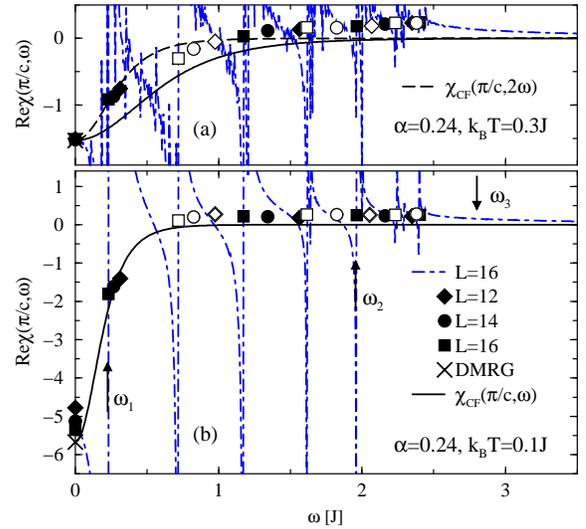}}
   \centerline{\parbox{\textwidth}{\caption{\label{Figure24}
   \sl Real part of the dimer-dimer correlation function for
   $\alpha=0.24$ at $k_{\rm B}T=0.1J$ (b) and $k_{\rm B}T=0.3J$
   (a). Dash-dotted lines: results applying Eq.\
   (\protect\ref{Rechispectral}). Symbols: values obtained via Eq.\  
   (\ref{Rechiregular}) with $\Delta\omega=0.1J$ for different system
   sizes. Full lines are the field-theoretical results from Eq.\
   (\ref{chiCF}), the dashed line is the scaled result, see Eq.\
   (\ref{Chiscale}).  
   }}}
   \end{figure}

At finite temperatures additional spectral lines attain significant 
weight as can be seen by the divergences in the real part of the
correlation function Eq.\ (\ref{Rechispectral}) as plotted for
$k_{\rm B}T=0.1J$ in Fig.\ \ref{Figure24} (b) and for $k_{\rm B}T=0.3J$
in Fig.\ \ref{Figure24} (a) (dash-dotted lines with L=16). The
most dominant and thus relevant spectral lines are $\tw_j \in
\tilde{\cal W}^{(16)}_{.24} = J\{0.233, 0.720, 1.174, 1.616, 1.963,
2.232, 2.403\}$. Again, the analogy to the unfrustrated Heisenberg
chain is obvious since $|\tilde{\cal W}^{(L)}_{.24}|=|\tilde{\cal
W}^{(L)}_{0}|$ for the chain lengths investigated.

\begin{figure}[bt]
   \epsfxsize=0.50\textwidth
   \centerline{\epsffile{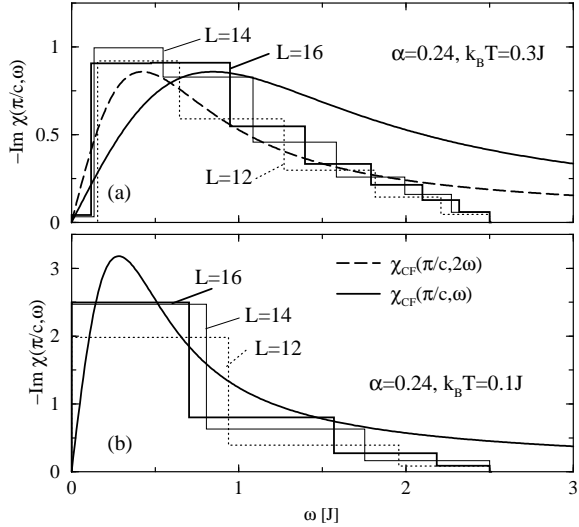}}
   \centerline{\parbox{\textwidth}{\caption{\label{FigIm24}
   \sl Imaginary part for $k_{\rm B}T=0.1J$ and $k_{\rm
   B}T=0.3J$ in the frustrated Heisenberg chain. Binned curves from
   Eq.\ (\protect\ref{Imchispectral}), dashed lines from field theory
   Eq.\ (\ref{chiCF}), the dashed line is rescaled as in Fig.\
   \protect\ref{Figure24}.   
   }}}
   \end{figure}

The imaginary part of the dimer-dimer correlation function is
determined for $k_{\rm B}T=0.1J$ via the same binning procedure as for $T=0$
and is shown in Fig.\ \ref{FigIm24} (b). For $k_{\rm B}T=0.3J$ the set of
frequencies given by $\tilde{\cal W}^{(16)}_{.24} \cup \{0\}$
enhancing the resolution as in the XY case. The resulting step
functions from the binning procedure Eq.\ (\ref{Imchibins}) are shown
in Fig.\ \ref{FigIm24} (a). In both cases the scaling regularization
Eq.\ (\ref{Imchidens}) is inappropriate as discussed in Sec.\
\ref{sectionregular}.

The symbols in Fig.\ \ref{Figure24} (a) and (b) show the
numerical results of the real part of the dimer-dimer correlation
function applying Eq.\ (\ref{Rechiregular}) for $k_{\rm B}T=0.3J$ and
$k_{\rm B}T=0.1J$, respectively, with $\Delta\omega=0.1J$. The full
symbols are at spectral lines ${\cal W}^{(L)}_{.24}$ and at
$\omega=0$, the open symbols are at $\tilde{\cal
W}^{(L)}_{.24}\setminus {\cal W}^{(L)}_{.24}$.  

For $\omega>2.5J$ the numerical results in Fig.\ \ref{Figure24} are
in the thermodynamic limit. The multitude of spectral
lines appearing  at finite temperatures which lead to the requirements
on $\Delta\omega$ can clearly be seen by the dash-dotted lines in
Fig.\ \ref{Figure24} (a) and (b). To obtain the smooth results of the
regularized data points a choice of $0.06J<\Delta\omega<0.15J$ is
necessary at intermediate temperatures $0.1J<k_{\rm B}T<0.9J$.

\begin{figure}[bt]
   \epsfxsize=0.50\textwidth
   \centerline{\epsffile{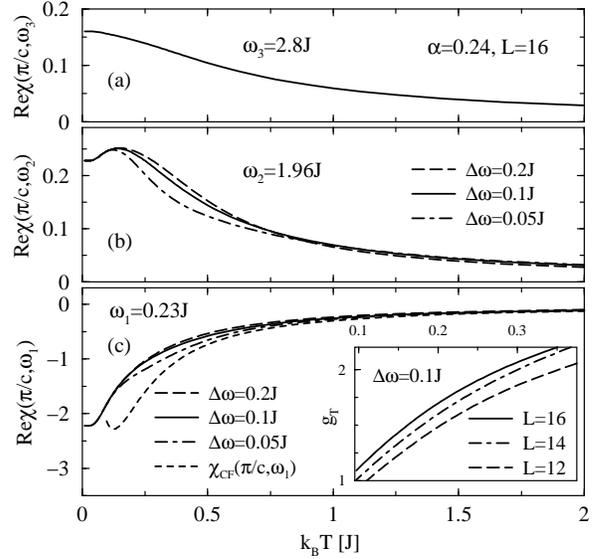}}
   \centerline{\parbox{\textwidth}{\caption{\label{ChivonT24}
   \sl Temperature dependence of ${\rm Re}\,\chi(\pi/c,\omega_i)$ for 
   $\omega_i$ as indicated in Fig.\ \protect\ref{Figure24}. (a)
   $\omega_3=2.8J$: result from Eq.\ 
   (\protect\ref{Rechispectral}). (b) $\omega_2=1.96J$ and (c)
   $\omega_1=0.23J$: results from Eq.\ (\ref{Rechiregular}), influence
   of $\Delta\omega$. For $\Delta\omega=0.05J$ the points in Fig.\
   \protect\ref{Figure24} scatter significantly at intermediate
   temperatures, the results are reliable for
   $0.06J<\Delta\omega<0.15J$. The short-dashed line in (c) is the
   field-theoretical result Eq.\ (\ref{chiCF}) deviating significantly
   for $k_{\rm B}T<0.6J$.    
   The inset shows the scaling function defined in Eq.\
   (\ref{Chiscale}) for different chain lengths. The frequency at
   $L=14$ is $\omega_1=0.27J$ and at $L=12$ it is $\omega_1=0.31J$. 
   }}}
   \end{figure}

The temperature dependent effect of the choice of $\Delta\omega$ is
shown in Fig.\ \ref{ChivonT24} for three different frequencies. At
$\omega_3=2.8J$ the result is equal to the thermodynamic limit. At
intermediate frequencies, as shown in the example of $\omega_2=1.96J$
and $\omega_1=0.23J$ in Fig.\ \ref{ChivonT24} (b) and (c),
respectively, values of $\Delta\omega<0.06J$ and $\Delta\omega>0.15J$
deviate from the best choice of $\Delta\omega<0.1J$. For other
temperatures, i.e., $0.1J>k_{\rm B}T$ and $k_{\rm B}T>0.9J$, one may
choose $\Delta\omega$ as small as $0.01J$. 
The influence of finite size effects can be estimated by comparing the 
$\omega=0$ values with DMRG results shown by crosses in Fig.\
\ref{Figure24}.\cite{Raupach99CGO} Together with the above discussed
small dependence of the values on $\Delta\omega$ this is strong
evidence for the accuracy of the results.

The field-theoretical result in Eq.\ (\ref{chiCF}) is expected to
well describe the low energy physics. The full lines in Fig.\
\ref{Figure24} show the corresponding curves. The functions are scaled
by $\chi_{\rbx{0}}(k_{\rm B}T/J)$ to match the $\omega=0$ values
obtained by DMRG.\cite{Raupach99CGO} While at $k_{\rm B}T=0.1J$ both the
imaginary and the real part of the dimer-dimer correlation function are
well described for $\omega\le J$ as shown in  Figs.\ \ref{FigIm24} and
\ref{Figure24} (b), at $k_{\rm B}T=0.3J$ the discrepancy to the
numerical result is obvious as shown in Figs.\ \ref{FigIm24} and
\ref{Figure24} (a). 

As in the unfrustrated case I attempt to improve the field-theoretical
result by introducing the scaling function $g_T$ defined in Eq.\
(\ref{Chiscale}). The relevant frequency in the frustrated case is
$\omega_1=0.23J$ for $L=16$. The temperature dependence of
$\chi_{\rbx{}}(\pi/c,\omega_1)$ is depicted by the full line in Fig.\
\ref{ChivonT24} (c) in comparison to field theory (short-dashed
line). The scaling function $g_T$ is shown in the inset of Fig.\
\ref{ChivonT24} (c). The dashed line in Fig.\ \ref{FigIm24} (a) shows
the rescaled imaginary part of the rescaled field theoretical curve,
the dashed line in Fig.\ \ref{Figure24} (a) shows the real part at
$k_{\rm B}T=0.3J$. Both fit the numerical data better than the
unscaled result (full lines). Above $k_{\rm B}T>0.3J$ the
field-theoretical function Eq.\ (\ref{chiCF}) is not a good fit
function any more since intermediate frequency values are
underestimated.  

In the temperature window $0.1J<k_{\rm B}T<0.3J$ relevant for
describing the quasi-elastic scattering in CuGeO$_3$ the above
described fitting procedure is sufficiently accurate. More precise
results for the correlation function can be obtained by fitting a
spline to the spectrum of the real part and simultaneaously fitting
its Kramers-Kronig transform to the imaginary part.

\section{Application to CuGeO$_3$}

CuGeO$_3$ undergoes a spin-Peierls transition at $T_{\rm SP}=14.3$ K or
$k_{\rm B}T_{\rm SP}\approx 0.1J$. The wave vector of the modulation in
the ordered phase is ${\bf q}_0=(\pi/a,0,\pi/c$), where $a$ and $c$
are the lattice constant along the crystallographic $x$ and $z$ direction,
respectively. I set $J/k_{\rm B}=150$ K which is together
with a value of $J_2/J=0.24$ among those discussed as valid for
CuGeO$_3$.\cite{Werner99CGO,Brenig97CGO}

\subsection{Quasi-elastic scattering}

Quasi-elastic scattering has been observed in X-ray scattering up to
40 K or $k_{\rm B}T\approx 0.3J$.\cite{Pouget96CGO} I can thus use the
results presented in Sec.\ \ref{sectionFrustrated} to compare with
experiment. In neutron scattering experiments the quasi-elastic
scattering is observable only up to 16
K.\cite{Hirota95CGO,BradenHabil}  

The frequency dependent scattering rate of inelastic neutrons or
X-rays can be obtained by convoluting the dynamical structure factor 
(\ref{dynamical}) with a Gaussian of the width of the experimental
resolution $\sigma$.
\begin{equation}\label{Intensity}
I({\bf q}_0,\omega)=\frac{1}{\sqrt{2\pi}\sigma}\int_{-\infty}^\infty
	{\rm e}^{-\frac{(\omega'-\omega)^2}{2\sigma^2}}\ 
        S({\bf q}_0,\omega')\ d\omega'
\end{equation}

The energy resolution in neutron scattering is of the order of a few
meV while X-rays integrate over a much larger energy interval. 
$\sigma=\sigma_n\approx0.05J$ simulates the resolution of
diffracted neutrons\cite{Gros98CGO} and $\sigma=\sigma_x\approx 0.5J$
is relevant for X-ray scattering. The X-ray resolution is probably
even larger, but the interval of $-0.5J<\omega<0.5J$ covers the full
width of the relevant magnetic spectrum, compare Fig.\
\ref{Figure24}. Also, the fit function Eq.\ (\ref{chiCF}) is not
reliable for $|\omega|>J$.

   \begin{figure}[bt]
   \epsfxsize=0.50\textwidth
   \centerline{\epsffile{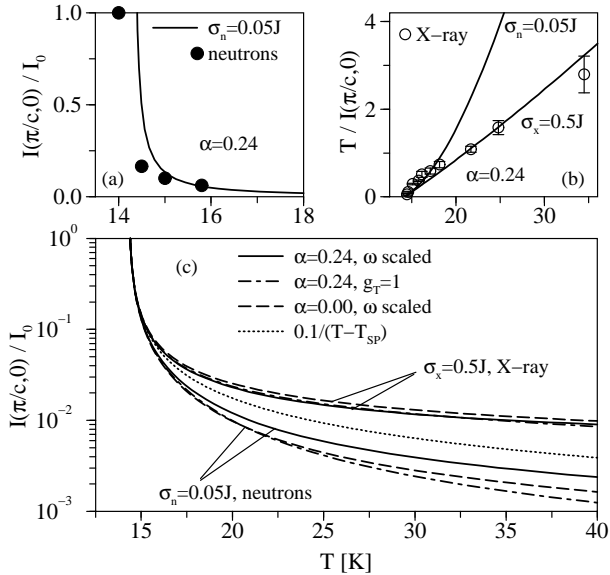}}
   \centerline{\parbox{\textwidth}{\caption{\label{FigIntensity}
   \sl Intensity of quasi-elastic scattering. (a) The fast drop
   $I\sim(T-T_{\rm SP})^{-1}$ above the phase transition at $T_{\rm
   SP}=14.3$ K is basically independent of the choice of parameters
   and matches the data from neutron scattering (filled circles, Ref.\
   \protect\onlinecite{BradenHabil}). (b) The curve of $T/I({\bf
   q}_0,0)$ obtained with the larger resolution $\sigma_x=0.5J$ match
   the X-ray data (open circles, scaled, Ref.\
   \protect\onlinecite{Pouget96CGO}) better than that obtained with 
   $\sigma_n=0.05J$. (c) The influence of the choice of parameters
   shown on a logarithmic scale. For $\sigma_x=0.5J$ the curves lie
   up to an order of magnitude higher than for $\sigma_n=0.05J$. The
   dot-dashed lines show the effect of omitting the frequency scaling
   ($g_T\equiv 1$). The dashed curves show the effect of the different
   functional dependence of $\chi_{\protect\rbx{0}}(k_{\rm B}T/J)$ in
   Eq.\ (\protect\ref{chiCF}) for $\alpha=J_2/J=0$.
   }}}
   \end{figure}

Fig.\ \ref{FigIntensity} (a) shows the fast drop of the intensity of
the quasi-elastic scattering above the phase transition. The
theoretical intensity (full line) is normalized with
$I_0=I({\bf q}_0,0)|_{T=14.4\,{\rm K}}$. It is basically independent of
the choice of parameters, all curves shown in Fig.\ \ref{FigIntensity}
(c) fall more or less on the same line on the linear scale. 

The full curve in Fig.\ \ref{FigIntensity} (a) matches the data from
neutron scattering (filled circles, Ref.\
\protect\onlinecite{BradenHabil}). Note that the critical region has
been estimated via the Ginzburg criterion\cite{Werner99CGO} to be
$T_{\rm SP}\pm 0.4$ K. Within this region the theoretical divergence
of the intensity $I=S({\bf q}_0,0)\sim (T-T_{\rm SP})^{-1}$ (dotted
line in Fig.\ \ref{FigIntensity} (c)) will be suppressed by critical
fluctuations. A more quantitative analysis also must include the limited
momentum resolution of the neutron data as discussed in Ref.\
\onlinecite{Holicki00XY}.

Fig.\ \ref{FigIntensity} (b) shows a plot of $T/I({\bf q}_0,0)$. As
desired, the curve obtained with the larger resolution $\sigma_x=0.5J$
match the X-ray data (open circles, scaled, Ref.\
\protect\onlinecite{Pouget96CGO}) better than that obtained with
$\sigma_n=0.05J$.

The influence of the choice of parameters becomes apparent on the
logarithmic scale shown in Fig.\ \ref{FigIntensity} (c). For
$\sigma_x=0.5J$ the curves lie an order of magnitude higher than for
$\sigma_n=0.05J$ at large temperatures. This explains why the
correlation length can be extracted to higher temperatures from X-ray 
measurements\cite{Pouget96CGO} then from neutron
scattering.\cite{Hirota95CGO} The dot-dashed lines show the effect of
omitting the frequency scaling ($g_T\equiv 1$). For $\sigma_n=0.05J$
the intensity is suppressed since the imaginary part without rescaling
has less weight in the low energy regime (see Fig.\ \ref{FigIm24}
(a)). For $\sigma_n=0.5J$ this effect is cancelled by the larger
contribution of the real part (Fig.\ \ref{Figure24} (a)). The dashed
curves show the effect of the different functional dependence of
$\chi_{\rbx{0}}(k_{\rm B}T/J)$ in Eq.\ (\protect\ref{chiCF}) for
$\alpha=J_2/J=0$.\cite{Raupach99CGO} 

Overall, the effect of energy resolution accounts for the largest
effects in the temperature dependence of the intensity of the
quasi-elastic scattering.  The effect of the frequency scaling $g_T$
is a minor correction. Both neutron and X-ray scattering intensities
are satisfactorily described within RPA.

\subsection{Phonon hardening}

The optical zone-boundary phonons coupling to the spin chains in
CuGeO$_3$ have been shown to harden with the lowering of
temperature.\cite{Braden98CGO} The phononic excitations in RPA are
essentially given by the roots of the real part of the denominator of
the normal coordinate propagator.\cite{Gros98CGO,Wer99,Holicki00XY}
\begin{equation}\label{roots}
{\rm Re}\,\chi({\bf q},\omega) =
   \left[\sum\nolimits_{\nu} 
         g_{\nu,{\bf q}}g_{\nu,-{\bf q}}\
                    D^{(0)}_{\nu}({\bf q},\omega)\right]^{-1}
\end{equation}
The coupling constants and bare phonon frequencies are given in Ref.\
\onlinecite{Werner99CGO}. The Peierls-active phonon mode coupling
strongest being $\Omega_{2,{\bf q}_0}=2.13J$, the left hand side of
Eq.\ (\ref{roots}) can be approximated by ${\rm Re}\,\chi({\bf
q},1.96J)$ as calculated in Sec.\ \ref{sectionFrustrated}. 

The resulting temperature evolution is up to a prefactor and a constant
given by Fig.\ \ref{ChivonT24} (b). In contrast to the case of
determining the dimer-dimer correlation function by the
field-theoretical result Eq.\ (\ref{chiCF}),\cite{Gros98CGO} the
phonon hardens with decreasing temperature while $k_{\rm
B}T>0.2J$. This is reminiscent of the change of sign of the real part
of the dimer-dimer correlation function discussed in Sec.\
\ref{sectionFrustrated}. The temperature evolution is in qualitative
agreement with the experimental data from Ref.\
\onlinecite{Braden98CGO}, but the phonon hardens only by 1\% which is
less than what is experimentally observed. The higher Peierls-active
phonon modes show no spin-phonon coupling induced temperature
dependence both in experiment and theory.  

\section{Conclusions}

Within the energy resolution limited by finite system sizes the
dimer-dimer correlation functions obtained by exact diagonalization
yield satisfactory approximations to the thermodynamic
limit. This is supported by the comparison with the exact results for
the XY model. The determination of both the real and the imaginary
part yields an essential improvement of the reliability of the results.

The quantitative applicability of field-theoretical results for the
dimer-dimer correlation functions is drawn into doubt, even at low
frequencies and temperatures. At intermediate temperatures they can be
used as fit functions by appropriately rescaling the amplitude and
frequency dependence.  

The intensity of the quasi-elastic scattering in CuGeO$_3$ is
satisfactorily reproduced within the RPA approach. The impact of
the energy resolution of the experimental setup on the temperature
dependence of the quasi-elastic scattering has been shown. The
numerical results yield the correct qualitative temperature dependence
of the Peierls-active phonons.

\section{Acknowledgments}

I thank K.\ Fabricius for calculating the matrix elements for system
size 16 and instructive discussions. I thank M.\ Karbach and G.\
M\"uller for the excitation data from Bethe ansatz and extensive
discussions on regularization procedures. I owe thanks to M.\ Braden
for furnishing the neutron scattering data files and R.\ Raupach for
the DMRG data files. I thank M.\ Braden, V.\ J.\ Emery, A.\ Kl\"umper,
S.\ Maslov, and R.\ Raupach for fruitful discussions.  The work
performed in Wuppertal was supported by DFG program ``Schwerpunkt
1073'', the work at BNL was supported by DOE contract number
DE-AC02-98CH10886.


\begin{thebibliography}{10}
\bibitem{Cutoffquote}
For a frustration of $\alpha=0.35$ the numerical analysis shows the cut off to
  be $\Lambda\approx 2.1J$. The effect of frustration is thus to compactify the
  dominant spectrum.
\end{thebibliography}
\end{document}